# Gradient Clock Synchronization using Reference Broadcasts

Fabian Kuhn      Rotem Oshman
fkuhn@csail.mit.edu   rotem@csail.mit.edu

Computer Science and Artificial Intelligence Laboratory,
Massachusetts Institute of Technology
Cambridge, MA 02139, USA

**Abstract**

In this paper we suggest a method by which reference broadcast synchronization (RBS), and other methods of estimating clock values, can be incorporated in standard clock synchronization algorithms to improve synchronization quality. We advocate a logical separation of the task of estimating the clock values of other nodes in the network from the task of using these estimates to output a logical clock value.

The separation is achieved by means of a virtual *estimate graph*, overlaid on top of the real network graph, which represents the information various nodes can obtain about each other. RBS estimates are represented in the estimate graph as edges between nodes at distance 2 from each other in the original network graph. A clock synchronization algorithm then operates on the estimate graph as though it were the original network.

To illustrate the merits of this approach, we modify a recent optimal gradient clock synchronization algorithm to work in this setting. The modified algorithm transparently takes advantage of RBS estimates and any other means by which nodes can estimate each others' clock values.

## 1   Introduction

The evolving field of wireless networks poses new and interesting challenges to time synchronization, leading to renewed attention to this venerable problem in recent years. Sensor networks in particular are subject to constraints on computation power and energy consumption, and often require a greater degree of synchronization than traditional distributed applications.

In a multi-hop sensor network it is frequently the case that neighboring nodes must be closely synchronized, while far-apart nodes can tolerate greater clock skew: neighboring nodes interfere with each other when they try to transmit, and are also more likely to cooperate for the purpose of some local computation. This gives rise to the problem of *gradient clock synchronization*, in which the synchronization between two nodes improves the closer they are to each other. The problem was first formulated in [6], where it is shown that in a network of diameter $D$, no algorithm can guarantee a skew that is better than $\Omega(\log D/ \log \log D)$ even between adjacent nodes. Subsequent work has improved the lower bound to $\Omega(\log D)$, and come up with algorithms that match it [11, 12].

The wireless broadcast medium also offers opportunities for better synchronization. Although contention may cause unpredictable delays before a message is broadcast, once a message is transmitted, it is received by all nodes in the sender's neighborhood almost instantaneously. Reference broadcast synchronization (RBS) [4] takes advantage of this to let the *neighbors* of the sender estimate each other's clock values with great accuracy. RBS can be extended to multi-hop networks, to allow any node in the network to estimate the clock value of any other node. However, by itself, RBS does not output a *logical clock* at every node, and so it is not a clock synchronization algorithm in the traditional sense.

In this paper we suggest an approach by which RBS, or any other estimation method (including external time sources), can be seamlessly incorporated in many clock synchronization algorithms, in order to reduce the effective diameter of the network and achieve better synchronization. We suggest a separation between the *estimate layer*, which is responsible for estimating other nodes' clock values, and the algorithm that uses these estimates to compute a local logical clock. The estimate layer runs underneath the algorithm and provides it with an *estimate graph* $G^{\text{est}}$. Each edge $\{u, v\}$ of $G^{\text{est}}$ represents an estimate that node $u$ can get for node $v$'s clock value (and vice-versa), along with an associated *uncertainty*. RBS estimates are represented in $G^{\text{est}}$ as edges between nodes at distance 2 from each other in the original network graph.

Almost any clock synchronization algorithm can be used on top of the estimate layer, as long as the algorithm can handle networks with non-uniform uncertainty on the links. The resulting synchronization between nodes $u, v$ depends on their *effective distance* $\text{dist}(u, v)$, and on the *effective diameter* of the network graph. These are defined by the corresponding distances in the estimate graph $G^{\text{est}}$. Using RBS it is possible to reduce the effective diameter to $O((\rho \cdot \mathcal{T} + u_{\text{rcv}}) \cdot D + \mathcal{T})$, where $D$ is the diameter of the original network, $\mathcal{T}$ is a bound on the message delay, $\rho$ is a bound on clock drift (typically very small), and $u_{\text{rcv}}$ is a bound on the receiver uncertainty (also very small [4]), which bounds the time it takes a node to process a message it receives.

Our main contributions are as follows. In Section 4 we define the estimate layer, and show how to incorporate point-to-point messages and RBS. In Section 5, we illustrate the applicability of our approach by modifying the algorithm of [12] to work on top of the estimate layer. Significantly, this involves extending it to a heterogeneous network; in [12] it is assumed that all links are subject to the same bounds on message delay. Finally, in Section 6 we prove that the algorithm achieves gradient clock synchronization, with the skew between nodes $u$ and $v$ bounded by $O(\text{dist}(u, v) \cdot \log_{1/\rho} D)$ in networks with effective diameter $D$ and drift bounded by $\rho$. This is asymptotically optimal. The proof is based on the proof in [12], but in our view it is cleaner and somewhat simpler.

In a companion paper to this one [10], we consider the problem of gradient clock synchronization in *dynamic* networks, and show that the weighted-graph approach employed here is useful in that context as well.

## 2  Related Work

The problem of establishing a common notion of time is at the core of many distributed systems and applications and has thus been widely studied, both from a theoretical and a practical point of view. In most of the existing work on clock synchronization, the nodes of a network compute estimates about each others' clock values by exchanging messages. Based on the obtained information, each node computes a local logical clock. Typically, the accuracy of clock estimates is determined by the uncertainty about the propagation delay of messages. In [13], it is shown that even if hardware clocks experience no drift, no clock synchronization algorithm can prevent a clock skew of $\Omega(D)$ in a network of diameter $D$. This lower bound on the maximum clock skew between any two nodes is matched by an algorithm described in [18] and by many subsequent algorithms (e.g. [2, 5, 12, 11, 14, 15]). Clock synchronization algorithms and lower bounds that accommodate non-uniform propagation delays are described, for example, in [1, 3, 7].

In [6], Fan and Lynch introduced the gradient clock synchronization problem. It is shown that even on a path of length $D$, no algorithm can guarantee a clock skew smaller than $\Omega(\log D/\log \log D)$ between adjacent nodes. This bound has been improved to $\Omega(\log D)$ in [12] and it is shown in [11, 12] that the new bound in indeed tight.

The special properties, constraints, and requirements of wireless ad hoc and sensor networks make clock synchronization especially challenging. There is a considerable amount of work on the problem (e.g. [5, 16, 17, 19]). Particularly interesting is the work on reference broadcast synchronization [4, 8], which exploits the property of sharing a single communication channel to obtain high accuracy clock estimates of nearby nodes.

## 3  Preliminaries

In the sequel we use $\mathbb{R}^{\geq 0}$ to denote the set of non-negative reals and $\mathbb{N}^{>0}$ to denote the positive integers.

We model a wireless network as an undirected graph $G = (V, E)$, where $V$ is the set of nodes, and $\{u, v\} \in E$ iff $u$ is in range of $v$ and vice-versa. We abstract away low-level details of contention management, message loss and so on, by assuming reliable message delivery with message delays bounded by a parameter $\mathcal{T}$.

Each node $v$ in the network has access to a local hardware clock $H_v$, which is subject to drift bounded by $\rho < 1$. We assume that for all $t_1 \leq t_2$,

$$(1 - \rho)(t_2 - t_1) \leq H_v(t_2) - H_v(t_1) \leq (1 + \rho)(t_2 - t_1).$$

It is also assumed that the hardware clock increases continuously and (for the analysis) is differentiable.

The goal of gradient clock synchronization is to output a local logical clock $L_v$ at every node $v$, which is closely-synchronized with all the other logical clocks. Formally, an algorithm is said to achieve $f$-gradient clock synchronization, for a function $f : \mathbb{R}^{\geq 0} \to \mathbb{R}^{\geq 0}$, if it satisfies the following requirement.

**Requirement 3.1.** *For all $u, v \in V$ and times $t$ we have*
$$L_v(t) - L_u(t) \leq f\left(\mathrm{dist}(u,v)\right).$$

Here $\mathrm{dist}(u,v)$ stands for the distance between $u$ and $v$; informally, the distance corresponds to the quality of information $u$ and $v$ can acquire about each other. Traditionally, $\mathrm{dist}(u,v)$ was defined to be the minimal sum of uncertainties about message delays on any path between $u$ and $v$. In this work we re-define $\mathrm{dist}(u,v)$ to take into account reference broadcast synchronization; for details, see Sec. 5.

In addition to $f$-gradient synchronization, we require the logical clocks to behave like a "real" clock. Specifically, the logical clocks should be non-decreasing, and they should always be within a linear envelope of real time. This is captured by the following requirement.

**Requirement 3.2.** *There exist $\alpha \in (0,1)$ and $\beta \geq 0$ such that for all $t_1 \leq t_2$,*
$$(1-\alpha)(t_2 - t_1) \leq L_u(t_2) - L_u(t_1) \leq (1+\beta)(t_2 - t_1).$$

In particular, the logical clocks are continuous. The algorithm we present in Sec. 5 outputs logical clocks that are also differentiable if the hardware clocks are differentiable.

## 4  The Estimate Layer

The estimate layer encapsulates point-to-point messages, reference broadcast synchronization, and any other means the nodes in the network have of obtaining information about the logical clock values of other nodes. The estimate layer provides an undirected *estimate graph* $G^{\mathrm{est}} = (V, E^{\mathrm{est}})$, where each edge $u, v \in E^{\mathrm{est}}$ represents some method by which nodes $u$ and $v$ can estimate each others' logical clock values. Note that $G^{\mathrm{est}}$ can be different from the underlying network graph $G$; for example, RBS is represented in $G^{\mathrm{est}}$ as edges connecting nodes at distance 2 from each other in $G$. We use $N(u)$ to denote $u$'s neighborhood in $G^{\mathrm{est}}$: $N(u) := \{v \in V \mid u, v \in E^{\mathrm{est}}\}$.

The estimate layer provides each node $u \in V$ with a set of local variables $\left\{\tilde{L}_u^v : v \in N(u)\right\}$, which represent $u$'s current estimates for the logical clock values of its neighbors in $G^{\mathrm{est}}$. Since the estimates are typically inaccurate, we associate with every edge $e \in E^{\mathrm{est}}$ an *uncertainty* $\epsilon_e$. The estimate layer guarantees the following property.

**Property 4.1** (Estimate quality). *For any edge $(u,v) \in E^{\mathrm{est}}$ and time $t$, we have*
$$L_v(t) - \epsilon_{\{u,v\}} \leq \tilde{L}_u^v(t) \leq L_v(t) + \epsilon_{\{u,v\}}.$$

Some common methods of obtaining logical clock estimates are described below. We describe each method and bound the error associated with it, and then show how to combine multiple methods so as to guarantee Property 4.1.

**Direct estimates.** As in [12], we assume that every node broadcasts its logical clock value to all its neighbors once every subjective $\Delta H$ time units (that is, after its hardware clock has increased by $\Delta H$), where $\Delta H$ is a parameter. These messages provide a direct estimate of the node's logical clock value. When a receive$(u, v, L)$ message occurs at time $t$, node $u$ sets

$$\tilde{L}_u^{v,\text{direct}} \leftarrow L.$$

Between messages from $v$, node $u$ increases $\tilde{L}_u^{v,\text{direct}}$ at the rate of its own hardware clock.

In Appendix A we show that the error of a direct estimate is bounded by

$$-(\alpha + \rho)\left(\frac{\Delta H}{1 - \rho} + \mathcal{T}\right) \leq L_v(t) - \tilde{L}_u^{v,\text{direct}}(t) \leq (\beta + \rho)\left(\frac{\Delta H}{1 - \rho} + \mathcal{T}\right) + (1 - \rho)\mathcal{T}.$$

(Note that at this point our error bound is asymmetric. We later show how to come up with a symmetric guarantee in the style of Prop. 4.1.)

**RBS estimates.** An RBS estimate is obtained by comparing the logical clock values that various nodes recorded when some common event occurred; in our case, the common event is a broadcast by a shared neighbor. We use $\mathcal{H}_u$ to denote node $u$'s *history*, a set of triplets $(x, H, L)$ where $x$ is a unique event identifier and $H, L$ are node $u$'s hardware and logical clock values (respectively) when it observed the event. After recording an event, the node sends a report$(u, x, L)$ message, which is propagated by other nodes until it reaches all other nodes that observed the same event. In our case, report$(\cdot)$ messages need to be re-broadcast only once, so that they reach the 2-neighborhood of the node that originated the report.

The accuracy of RBS depends on two factors.

1. Receiver uncertainty: this is the time required for nodes to process the common event and record their logical clock value. The receiver uncertainty is bounded by $u_{\text{rcv}}$ if whenever an event $x$ occurs at real time $t$, there is some $L \in [L_u(t), L_u(t + u_{\text{rcv}})]$ such that for all $t' \geq t + u_{\text{rcv}}$ we have $(x, L) \in \mathcal{H}_u(t')$.

2. Propagation delay: report$(\cdot)$ messages are subject to the usual message delay. This contributes to the inaccuracy of the estimate, because while the report is propagated the clocks may continue to drift apart.

   We say that the propagation delay is bounded by $\mathcal{P}$ if whenever a node $u$ experiences an event $x$ at real time $t$, every node $v \in N^2(u)$ receives a report$(u, x, L)$ message no later than time $t + \mathcal{P}$.

   In our case, because report$(\cdot)$ messages need to be re-broadcast only once, the propagation delay is bounded by $\mathcal{P} \leq u_{\text{rcv}} + 2\left(\frac{\Delta H}{1-\rho} + \mathcal{T}\right)$: after observing the event, node $u$ waits at most $\frac{\Delta H}{1-\rho}$ time units and then broadcasts the message, which takes at most $\mathcal{T}$ time units to arrive; its neighbors do the same.

When node $u$ receives a report$(v, x, L)$ message at time $t$, it looks up the corresponding triplet $(x, H', L')$ recorded in its own history. It uses $H_u - H'$ to estimate the time that has passed since $x$ occurred, and sets
$$\tilde{L}_u^{v,\text{rbs}} \leftarrow L + H_u - H'.$$
Every broadcast by a node is an event that its neighbors can use to get estimates of each others' logical clock values.

In Appendix A we show that RBS estimates are accurate up to the following bound.

$$-(\alpha + \rho)\left(\frac{\Delta H}{1-\rho} + \mathcal{P}\right) - (1-\alpha)u_{\text{rcv}} \leq L_v(t) - \tilde{L}_u^{v,\text{rbs}}(t) \leq$$
$$\leq (\beta + \rho)\left(\frac{\Delta H}{1-\rho} + \mathcal{P}\right) + (1-\rho)u_{\text{rcv}}.$$

**Combining multiple estimates.** As we have seen, each node may have multiple ways of estimating the clock values of its neighbors in $G^{\text{est}}$. Let $\tilde{L}_u^{v,1}, \ldots, \tilde{L}_u^{v,m}$ be the various estimates that $u$ has for $v$'s logical clock value, and let $\epsilon_{\text{low}}^1, \ldots, \epsilon_{\text{low}}^m$ and $\epsilon_{\text{high}}^1, \ldots, \epsilon_{\text{high}}^m$ be error bounds such that for all $i \in \{1, \ldots, m\}$ and time $t$,

$$-\epsilon_{\text{low}}^i \leq L_v(t) - \tilde{L}_u^{v,i}(t) \leq \epsilon_{\text{high}}^i. \tag{1}$$

Node $u$ computes a combined estimate with symmetric error, given by

$$\tilde{L}_u^v(t) := \frac{\min_i\left(\tilde{L}_u^{v,i}(t) + \epsilon_{\text{high}}^i\right) - \max_i\left(\tilde{L}_u^{v,i}(t) - \epsilon_{\text{low}}^i\right)}{2}. \tag{2}$$

The uncertainty of the combined estimate is bounded by

$$\epsilon_{\{u,v\}} := \min_i \left\{\frac{\epsilon_{\text{low}}^i + \epsilon_{\text{high}}^i}{2}\right\}.$$

## 5 An Optimal Gradient Clock-Synchronization Algorithm

In this section we modify the algorithm of [12] to work on top of the estimation layer presented in the previous section. To satisfy Requirement 3.2, the algorithm increases the logical clock in a continuous manner, with no discrete jumps. At each point during the execution a node is either in *fast mode* or in *slow mode*. In slow mode, $u$ increases its logical clock at a rate of $\frac{d}{dt}H_u(t)$; in fast mode, the logical clock rate is $(1+\mu)\frac{d}{dt}H_u(t)$, where $\mu$ is a parameter.

Each node continually examines its estimates for the logical clock values of its neighbors in $G^{\text{est}}$. To compensate for the uncertainty on edge $e$ we use a parameter $\kappa_e$, which must satisfy the following.

**Property 5.1** (Requirement on $\kappa_e$). *For any edge $e \in E$ we require $\kappa_e > \frac{1}{\lambda} \cdot \epsilon_e$, where $\lambda < \frac{1}{4}$ is some constant.*

If a node $u$ finds that it is too far behind, it goes into fast mode and uses the fast rate of $(1+\mu)\frac{d}{dt}H_u(t)$. The following rule is used to determine when to go into fast mode; informally, it states that some neighbor is far ahead, and no neighbor is too far behind.

**Definition 5.1** (Fast condition FC). *At time $t$, a node $u \in V$ satisfies the fast condition, denoted FC, if there is some integer $s \in \mathbb{N}$ for which following conditions are satisfied:*

(FC1) $\exists v \in N(u) : \tilde{L}_u^v(t) - L_u(t) \geq (s - 1 - \lambda)\,\kappa_{\{u,v\}}$, *and*

(FC2) $\forall v \in N(u) : L_u(t) - \tilde{L}_u^v(t) \geq (s - 1 + \lambda)\,\kappa_{\{u,v\}}$.

Conversely, if a node is far behind some neighbor, and no other neighbor is too far ahead of it, it enters slow mode and uses the slow rate. The following rule is used to determine when to enter slow mode.

**Definition 5.2** (Slow condition SC). *At time $t$, a node $u \in V$ satisfies the slow condition, denoted SC, if there is an integer $s \in \mathbb{N}^{>0}$ for which the following conditions are satisfied:*

(SC1) $\exists v \in N(u) : L_u(t) - \tilde{L}_u^v(t) \geq \left(s - \frac{1}{2} - \lambda\right) \cdot \kappa_{\{u,v\}}$, *and*

(SC2) $\forall v \in N(u) : \tilde{L}_u^v(t) - L_u(t) \leq \left(s - \frac{1}{2} + \lambda\right) \cdot \kappa_{\{u,v\}}$.

To show that the algorithm is realizable, we show that the two conditions are disjoint.

**Lemma 5.2.** *No node can satisfy SC and FC at the same time.*

*Proof.* Suppose by way of contradiction that $u$ satisfies both SC and FC at time $t$. Then there is an integer $s \in \mathbb{N}^{>0}$ for which

(SC1) $\exists v \in N(u) : L_u(t) - \tilde{L}_u^v(t) \geq \left(s - \frac{1}{2} - \lambda\right) \cdot \kappa_{\{u,v\}}$, and

(SC2) $\forall v \in N(u) : \tilde{L}_u^v(t) - L_u(t) \leq \left(s - \frac{1}{2} + \lambda\right) \cdot \kappa_{\{u,v\}}$,

and there is another integer $s' \in \mathbb{N}^{>0}$ such that

(FC1') $\exists v \in N(u) : \tilde{L}_u^v(t) - L_u(t) \geq (s' - 1 - \lambda) \cdot \kappa_{\{u,v\}}$, and

(FC2') $\forall v \in N(u) : L_u(t) - \tilde{L}_u^v(t) \leq (s' - 1 + \lambda) \cdot \kappa_{\{u,v\}}$.

For the node $v$ whose existence is stipulated in condition (FC1'), we have

$$\left(s' - 1 - \lambda\right) \cdot \kappa_{\{u,v\}} \stackrel{\text{(FC1')}}{\leq} \tilde{L}_u^v(t) - L_u(t) \stackrel{\text{(SC2)}}{\leq} \left(s - \frac{1}{2} + \lambda\right) \cdot \kappa_{\{u,v\}},$$

which implies that $s' \leq s + \frac{1}{2} + 2\lambda$. However, from condition (SC1), there exists a node $v'$ for which

$$\left(s - \frac{1}{2} - \lambda\right) \cdot \kappa_{\{u,v\}} \stackrel{\text{(SC1)}}{\leq} L_u(t) - \tilde{L}_u^{v'}(t) \stackrel{\text{(FC2')}}{\leq} \left(s' - 1 + \lambda\right) \cdot \kappa_{\{u,v\}},$$

and hence $s' \geq s + \frac{1}{2} - 2\lambda$. Since $\lambda < \frac{1}{4}$, together we have $s < s' < s + 1$, which is impossible because $s$ and $s'$ are integers. $\square$

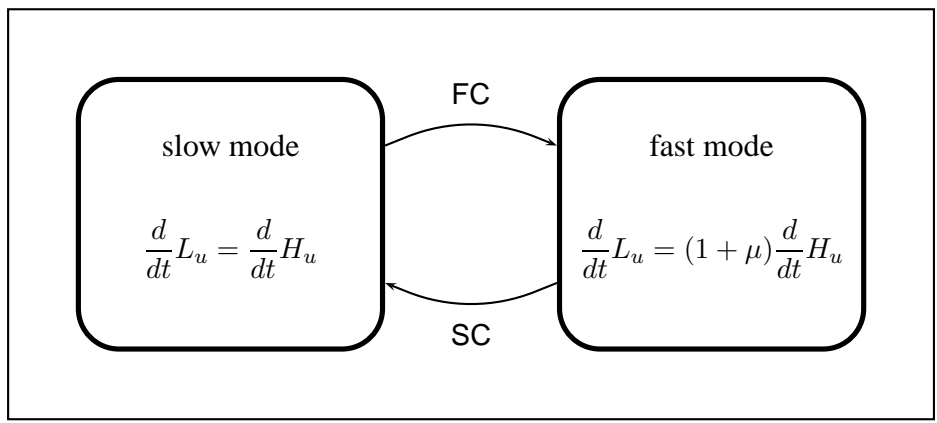

Figure 1: The automaton from Alg. 1

A formal description of the algorithm, in the form of a timed I/O automaton (see [9]), is given in Alg. 1. The switching between the modes is depicted in Figure 1.

## 6 Analysis

We define a parameter $\sigma \geq 2$, which serves as the base for the logarithm in the gradient skew bound. The correctness of the algorithm relies on the following assumption, which (informally) states that $\mu$ is large enough to allow nodes that are behind to catch up.

**Property 6.1** (Requirement on $\mu$). *We require*

$$\mu > 4\sigma \frac{\rho}{1 - \rho}. \tag{3}$$

```
automaton ClockSync
    signature
        internal enter_fast_mode, enter_slow_mode
    variables
        $H_u$ : Real := 0
        $L_u$ : Real := 0
        $\tilde{L}_u^v$ : Real := 0, for each $v \in N(u)$
        $\mathrm{mult}_u$ : discrete Real := 1
        $\mathrm{mode}_u$ : Boolean := slow
    transitions
        internal enter_fast_mode
            precondition:
                FC

            effect:
                $\mathrm{mode}_u$ := fast
                $\mathrm{mult} := 1 + \mu$

        internal enter_slow_mode
            precondition:
                SC

            effect:
                $\mathrm{mode}_u$ := slow
                $\mathrm{mult} := 1$

    trajectories
        stop when
            (SC holds and $\mathrm{mode}_u$ = fast) or (FC holds and $\mathrm{mode}_u$ = slow)

        evolve
            $1 - \rho \leq \frac{d}{dt} H_u \leq 1 + \rho$
            $\frac{d}{dt} L_u = \mathrm{mult}_u \cdot \frac{d}{dt} H_u$
```

**Algorithm 1**: A TIOA formulation for the algorithm from Section 5

Let $\mathcal{P}$ denote the set of all paths in the graph $G^{\text{est}}$ (including non-simple paths), and let $\mathcal{P}(v) \subseteq \mathcal{P}$ denote the set of paths that start at node $v$. Given a path $P = v_0, \ldots, v_k \in \mathcal{P}$, we denote $\kappa_P := \sum_{i=0}^{k-1} \kappa_{\{v_i, v_{i+1}\}}$. Given two nodes $u, v \in V$, the distance between $u$ and $v$ is defined by

$$\text{dist}(u,v) := \min_{P=u,\ldots,v} \kappa_P, \tag{4}$$

and the diameter of the graph $G^{\text{est}}$ is defined by

$$\mathcal{D} := \max_{u,v} \text{dist}(u,v). \tag{5}$$

We show that the following invariant, which we denote $\mathcal{L}$, is maintained throughout any execution of the algorithm.

**Definition 6.1** (Legal State). *We say that the network is in a* legal state *at time $t$ if and only if for all $s \in \mathbb{N}^{>0}$ and all paths $P = v_0, \ldots, v_k$, if*

$$\kappa_P(t) \geq C_s := \frac{2\mathcal{D}}{\sigma^s},$$

*then*

$$L_{v_k}(t) - L_{v_0}(t) \leq s \cdot \kappa_P.$$

In particular, if the network is legal at time $t$, then for every two nodes $u, v$ and integer $s \geq 1$ such that $\text{dist}(u,v) \geq C_s$, we have $L_u(t) - L_v(t) \leq s \cdot \text{dist}(u,v)$.

To show that the network is always in the safety region defined by the legal state condition, we show that whenever some path comes close to having illegal skew, the algorithm acts to decrease the skew, pulling the system back into the safety region.

We cannot guarantee that a node will always "realize" when it is on a path that has too much skew: each node only has knowledge of its local neighborhood, and this local image may not reflect a large skew further down the path. We can, however, show that when the skew is close to being illegal, the nodes that are "the most behind" or "the most ahead" (in the sense defined formally below) *do* realize that they must act to correct the skew. We will show that such nodes enter fast or slow mode as appropriate.

Since we can only argue about nodes that, roughly speaking, maximize some notion of weighted skew (defined below), we will employ the following technical lemma several times.

**Lemma 6.2.** *Let $g_1, \ldots, g_n : \mathbb{R}^{\geq 0} \to \mathbb{R}^{\geq 0}$ be differentiable functions, and let $[a,b]$ be an interval such that for all $i \in \{1, \ldots, n\}$ and $x \in (a,b)$, if $g_i(x) = \max_j g_j(x)$ then $\frac{d}{dx} g_i(x) \leq r$. Then for all $x \in [a,b]$, $\max_i g_i(x) \leq \max_i g_i(a) + r \cdot (x-a)$.*

We define two different notions of "weighted skew": one captures how much a node $v_0$ is ahead of any other node, and the other captures how far behind it is. The weights in both cases are proportional to the uncertainty on the path, but use different constants. These notions correspond exactly to the the fast and slow conditions, respectively.

**Definition 6.2.** *Given an integer $s \in \mathbb{N}$, a time t, and a path $P = v_0, \ldots, v_k \in \mathcal{P}$, we define*

$$\Xi_P^s(t) := L_{v_0}(t) - L_{v_k}(t) - (s-1) \cdot \kappa_P, \quad \text{and} \quad \Xi_{v_0}^s(t) := \max_{P \in \mathcal{P}(v_0)} \Xi_P^s(t).$$

**Definition 6.3.** *Given an integer $s \in \mathbb{N}$, a time t, and a path $P = v_0, \ldots, v_k \in \mathcal{P}$, we define*

$$\Psi_P^s(t) := L_{v_k}(t) - L_{v_0}(t) - \left(s - \frac{1}{2}\right) \cdot \kappa_P, \quad \text{and} \quad \Psi_{v_0}^s(t) := \max_{P \in \mathcal{P}(v_0)} \Psi_P^s(t).$$

These definitions induce "inner safety regions" $\mathbf{\Xi}^s := [\max_v \Xi_v^s \leq 0]$ and $\mathbf{\Psi}^s := [\max_v \Psi_v^s \leq 0]$ for any $s \in \mathbb{N}^{>0}$, with $\mathbf{\Xi}^s \subseteq \mathbf{\Psi}^s \subseteq \mathcal{L}$ (see Fig. 2).

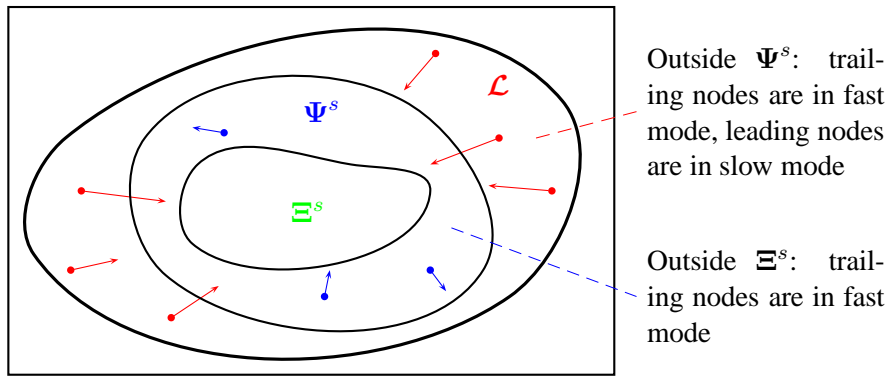

Figure 2: Regions $\mathbf{\Xi}^s$, $\mathbf{\Psi}^s$ and $\mathcal{L}$. Arrows illustrate the possible dynamics acting on the weighted skew in each region.

The next lemma can be thought of as bounding how far the system can stray outside the boundary of $\mathbf{\Xi}^s$ and $\mathbf{\Psi}^s$ while still being in a legal state.

**Lemma 6.3.** *If the network is in a legal state at time t, then for all nodes $u \in V$ and integers $s \geq 1$ we have $\Xi_u^s(t) < C_{s-1}$  $\Psi_u^s(t) < C_{s-1}$.*

*Proof.* It is sufficient to show that for all paths $P = u, \ldots, v$ and for all $s \geq 1$,

$$L_u(t) - L_v(t) - (s-1) \cdot \kappa_P \stackrel{?}{<} C_{s-1}. \tag{6}$$

Let $r \geq 1$ be the minimal integer such that $\text{dist}(u,v) \geq C_r$. (Note that $\text{dist}(u,v) \leq \mathcal{D} < C_0$, and therefore $r$ is well-defined.) Because the system is in a legal state at time $t$, we have

$$L_u(t) - L_v(t) \leq r \cdot \text{dist}(u,v) \leq r \cdot \kappa_P,$$

which can be re-written as

$$L_u(t) - L_v(t) - (s-1) \cdot \kappa_P \leq (r - s + 1) \cdot \kappa_P.$$

It is therefore sufficient to show that

$$(r - s + 1) \cdot \kappa_P \stackrel{?}{<} C_{s-1}. \tag{7}$$

If $r < s$, the left-hand side of (7) is at most 0. Since $C_{s-1} > 0$, (7) holds in this case. Otherwise, if $r \geq s$, then by choice of $r$ we have $\kappa_P < C_{r-1}$, and therefore,

$$(r - s + 1) \cdot \kappa_P < (r - s + 1) \cdot C_{r-1} = \frac{r - s + 1}{\sigma^{r-s}} \cdot C_{s-1} \leq C_{s-1}.$$

The last inequality holds because $x + 1 \leq 2^x$ for all $x \in \mathbb{N}$ and $\sigma \geq 2$. This shows that (7) holds in this case as well. □

Next we show that while the system is outside the region $\Xi^s$, nodes that are "the most behind" (maximize $\Xi$ with respect to some other node) will be acting to catch up, and while the system is outside the region $\Psi^s$, nodes that are "the most ahead" (maximize $\Psi$ with respect to some other node) will be held back from moving too quickly.

**Lemma 6.4.** *Given $s \in \mathbb{N}$, a node $v_0 \in V$, and a time $t$, let $P = v_0, \ldots, v_k \in \mathcal{P}(v_0)$ be a path starting at $v_0$ for which $\Xi_P^s(t) = \Xi_{v_0}^s(t)$. If $\Xi_{v_0}^s(t) > 0$, then $v_k$ is in fast mode.*

**Lemma 6.5.** *Given $s \in \mathbb{N}$, a node $v_0 \in V$, and a time $t$, let $P = v_0, \ldots, v_k \in \mathcal{P}(v_0)(t)$ be a path starting at $v_0$ for which $\Psi_P^s(t) = \Psi_{v_0}^s(t)$. If $\Psi_{v_0}^s(t) > 0$, then $v_k$ is in slow mode.*

The proofs of the two lemmas are very similar.

*Proof of Lemma 6.4.* We set out to show that $v_k$ satisfies FC.

Consider any path $P' = v_0, \ldots, v \in \mathcal{P}(v_0)$ that ends at a neighbor $v$ of $v_k$. Since $\Xi_P^s(t) = \Xi_{v_0}^s(t) = \max_{Q \in \mathcal{P}(v_0)} \Xi_Q^s(t)$, we have $\Xi_{P'}^s(t) \leq \Xi_P^s(t)$; that is,

$$L_{v_0}(t) - L_v(t) - (s - 1) \cdot \kappa_{P'} \leq L_{v_0}(t) - L_{v_k}(t) - (s - 1) \cdot \kappa_P.$$

Re-arranging yields $L_v(t) - L_{v_k}(t) \geq (s - 1) \cdot (\kappa_P - \kappa_{P'})$, and applying Property 4.1 we obtain

$$\tilde{L}_{v_k}^v(t) - L_{v_k}(t) \geq L_v(t) - \epsilon_{\{v,v_k\}} - L_{v_k}(t) \geq$$
$$\geq (s - 1) \cdot (\kappa_P - \kappa_{P'}) - \epsilon_{\{v,v_k\}}. \tag{8}$$

To show (FC1) is satisfied, let $P'$ be the subpath $v_0, \ldots, v_{k-1}$ of $P$, where $v_{k-1} \in N(v)$. Note that since $\Xi_P(t) > 0$ it must be that $k > 0$, and thus $v_{k-1}$ is well-defined. For this choice of $P'$, (8) yields

$$\tilde{L}_{v_k}^{v_{k-1}}(t) - L_{v_k}(t) \geq (s - 1) \cdot (\kappa_P - \kappa_{P'}) - \epsilon_{\{v_{k-1}, v_k\}} =$$
$$= (s - 1) \cdot \kappa_{\{v_{k-1}, v_k\}} - \epsilon_{\{v_{k-1}, v_k\}} \stackrel{\text{(Prop. 5.1)}}{>} (s - 1 - \lambda) \kappa_{\{v_{k-1}, v_k\}}.$$

This shows that (FC1) is satisfied. To show that (FC2) holds, let $v \in N(v_k)$ be any neighbor of $v_k$, and let $P' = v_0, \ldots, v_k, v$ be the path obtained by appending $v$ to the path $P$. In this case (8) yields

$$L_{v_k}(t) - \tilde{L}^v_{v_k}(t) \le (s-1) \cdot (\kappa_{P'} - \kappa_P) + \epsilon_{\{v,v_k\}} =$$
$$= (s-1) \cdot \kappa_{\{v,v_k\}} + \epsilon_{\{v,v_k\}} \overset{\text{(Prop. 5.1)}}{<} (s-1+\lambda) \cdot \kappa_{\{v,v_k\}}.$$

Hence, the second condition is satisfied as well, and node $v_k$ is in fast mode. □

*Proof of Lemma 6.5.* We set out to show that $v_k$ satisfies SC.

Consider any path $P' = v_0, \ldots, v \in \mathcal{P}(v_0)$ that ends at a neighbor $v$ of $v_k$. As before, since $\Psi^s_P(t) = \Psi^s_{v_0}(t) = \max_{Q \in \mathcal{P}(v_0)} \Psi^s_Q(t)$, we can write

$$L_v(t) - L_{v_0}(t) - \left(s - \frac{1}{2}\right) \cdot \kappa_{P'} \le L_{v_k}(t) - L_{v_0}(t) - \left(s - \frac{1}{2}\right) \cdot \kappa_P, \tag{9}$$

and again applying Property 4.1 we obtain

$$\tilde{L}^v_{v_k}(t) - L_{v_k}(t) \le L_v(t) + \epsilon_{\{v,v_k\}} - L_{v_k}(t) \le$$
$$\le \left(s - \frac{1}{2}\right) \cdot (\kappa_{P'} - \kappa_P) + \epsilon_{\{v,v_k\}}. \tag{10}$$

For (SC1), consider the subpath $P' = v_0, \ldots, v_{k-1}$ of $P$, where $v_{k-1} \in N(v_k)$. Inequality (10) yields

$$L_{v_k}(t) - \tilde{L}^{v_{k-1}}_{v_k}(t) \ge \left(s - \frac{1}{2}\right) \cdot (\kappa_P - \kappa_{P'}) - \epsilon_{\{v_{k-1},v_k\}} =$$
$$= \left(s - \frac{1}{2}\right) \cdot \kappa_{\{v_{k-1},v_k\}} - \epsilon_{\{v_{k-1},v_k\}} >$$
$$\overset{\text{(Prop. 5.1)}}{>} \left(s - \frac{1}{2} - \lambda\right) \cdot \kappa_{\{v_{k-1},v_k\}},$$

and so (SC1) is satisfied. For (SC2), let $v \in N(v_k)$ be any neighbor of $v_k$, and let $P' = v_0, \ldots, v_k, v$ be the path obtained by appending $v$ to $P$. Inequality (10) now gives

$$\tilde{L}^v_{v_k}(t) - L_{v_k}(t) \le \left(s - \frac{1}{2}\right) \cdot (\kappa_{P'} - \kappa_P) + \epsilon_{\{v,v_k\}} =$$
$$= \left(s - \frac{1}{2}\right) \cdot \kappa_{\{v,v_k\}} + \epsilon_{\{v,v_k\}} <$$
$$\overset{\text{(Prop. 5.1)}}{<} \left(s - \frac{1}{2} + \lambda\right) \cdot \kappa_{\{v,v_k\}},$$

which shows that (SC2) is satisfied as well. □

Suppose that at time $t$, node $v$ has $\Xi_v^s(t) > 0$. From Lemma 6.4, all the nodes that maximize $\Xi_v^s$ are in fast mode, trying to catch up to $v$, and their logical clock rate is at least $(1-\rho)(1+\mu)$. Thus, whenever it is positive, $\Xi_v^s$ decreases at an average rate of at least $(1-\rho)(1+\mu)$, *minus* the rate by which $v$ increases its own logical clock. To formalize this observation, define

$$\mathcal{I}_v(t_1, t_2) := L_v(t_2) - L_v(t_1) \tag{11}$$

to be the amount by which $v$ increases its logical clock over the time interval $[t_1, t_2]$. Since $\frac{d}{dt} L_v(t) \geq \frac{d}{dt} H_v(t) \geq 1 - \rho$ we have the following property.

**Property 6.6.** *For all nodes $v$ and times $t_1, t_2$ we have $\mathcal{I}_v(t_1, t_2) \geq (1-\rho)(t_2 - t_1)$.*

Now we can state the following lemma.

**Lemma 6.7** (Catch-Up Lemma). *Let $v_0$ be a node and let $[t_0, t_1]$ be a time interval such that for all $t \in (t_0, t_1)$ we have $\Xi_{v_0}^s(t) > 0$. Then for all $t \in [t_0, t_1]$,*

$$\Xi_{v_0}^s(t) \leq \Xi_{v_0}^s(t_0) + \mathcal{I}_{v_0}(t_0, t) - (1-\rho)(1+\mu)(t - t_0). \tag{12}$$

Similarly, whenever $\Psi_v^s(t) > 0$, the nodes that maximize $\Psi_v^s$ are in slow mode, and their logical clocks increase at a rate of at most $1+\rho$. Thus, whenever it is positive, $\Psi_v^s(t)$ increases at an average rate of at most $1 + \rho$, again minus $v$'s increase to its own logical clock. This is captured by the following lemma.

**Lemma 6.8** (Waiting Lemma). *Let $v_0$ be a node and let $[t_0, t_1]$ be a time interval such that for all $t \in (t_0, t_1)$ we have $\Psi_{v_0}^s(t) > 0$. Then for all $t \in [t_0, t_1]$,*

$$\Psi_{v_0}^s(t) \leq \Psi_{v_0}^s(t_0) - \mathcal{I}_{v_0}(t_0, t) + (1+\rho)(t - t_0). \tag{13}$$

The proofs of Lemmas 6.7 and 6.8 involve a straightforward application of Lemma 6.2.

*Proof of Lemma 6.7.* Consider the set of functions $\{g_P\}_{P \in \mathcal{P}(v_0)}$ defined by $g_P(t) := \Xi_P^s(t) - \mathcal{I}_{v_0}(t_0, t)$. Observe that for all $t$,

$$\max_{P \in \mathcal{P}(v_0)} g_P(t) = \max_P \left( \Xi_P^s(t) - \mathcal{I}_{v_0}(t_0, t) \right) =$$

$$= \left( \max_P \Xi_P^s(t) \right) - \mathcal{I}_{v_0}(t_0, t) =$$

$$= \Xi_{v_0}^s(t) - \mathcal{I}_{v_0}(t_0, t). \tag{14}$$

In addition, $\mathcal{I}_{v_0}(t_0, t_0) = 0$. Therefore (12) can be re-written as

$$\max_{P \in \mathcal{P}(v_0)} g_P(t) \stackrel{?}{\leq} \max_{P \in \mathcal{P}(v_0)} g_P(t_0) - (1-\rho)(1+\mu)(t - t_0). \tag{15}$$

Next, substituting the definition for $\Xi_{v_0}^s(t)$ and $\mathcal{I}_{v_0}(t_0, t)$ we obtain

$$g_P(t) = L_{v_0}(t) - L_{v_k}(t) - (s-1) \cdot \kappa_P - L_{v_0}(t) + L_{v_0}(t_0) =$$
$$= -L_{v_k}(t) - (s-1) \cdot \kappa_P - L_{v_0}(t_0),$$

and therefore,

$$\frac{d}{dt} g_P(t) = -\frac{d}{dt} L_{v_k}(t). \tag{16}$$

If $P$ is a path such that $g_P(t) = \max_Q g_Q(t)$, then $P$ also has $\Xi_P^s(t) = \Xi_{v_0}^s(t)$. Since $\Xi_{v_0}^s(t) > 0$ for any $t \in (t_0, t_1)$, we can apply Lemma 6.4 to obtain that whenever $g_P(t) = \max_Q g_Q(t)$ where $P = v_0, \ldots, v_k$, node $v_k$ is in fast mode and $\frac{d}{dt} L_{v_k}(t) = (1+\mu)\frac{d}{dt} H_{v_k}(t) \geq (1-\rho)(1+\mu)$. This is sufficient to apply Lemma 6.2 to the interval $[t_0, t]$, which yields (15). $\square$

*Proof of Lemma 6.8.* Consider the set of functions $\{g_P\}_{P \in \mathcal{P}(v_0)}$ defined by $g_P(t) := \Psi_P^s(t) + \mathcal{I}_{v_0}(t_0, t)$. Observe that for all $t$,

$$\max_{P \in \mathcal{P}(v_0)} g_P(t) = \max_P \left(\Psi_P^s(t) + \mathcal{I}_{v_0}(t_0, t)\right) =$$
$$= \left(\max_P \Psi_P^s(t)\right) + \mathcal{I}_{v_0}(t_0, t) =$$
$$= \Psi_{v_0}^s(t) + \mathcal{I}_{v_0}(t_0, t). \tag{17}$$

In addition, $\mathcal{I}_{v_0}(t_0, t_0) = 0$. Therefore (13) can be re-written as

$$\max_{P \in \mathcal{P}(v_0)} g_P(t) \stackrel{?}{\leq} \max_{P \in \mathcal{P}(v_0)} g_P(t_0) + (1+\rho)(t - t_0). \tag{18}$$

Next, substituting the definitions for $\Psi_{v_0}^s(t)$ and $\mathcal{I}_{v_0}(t_0, t)$ we obtain

$$g_P(t) = L_{v_k}(t) - L_{v_0}(t) - \left(s - \frac{1}{2}\right) \cdot \kappa_P + L_{v_0}(t) - L_{v_0}(t_0) =$$
$$= L_{v_k}(t) - \left(s - \frac{1}{2}\right) \cdot \kappa_P - L_{v_0}(t_0),$$

and therefore,

$$\frac{d}{dt} g_P(t) = \frac{d}{dt} L_{v_k}(t). \tag{19}$$

If $P$ is a path such that $g_P(t) = \max_Q g_Q(t)$, then $P$ also has $\Psi_P^s(t) = \Psi_{v_0}^s(t)$. Since $\Psi_{v_0}^s(t) > 0$ for any $t \in (t_0, t_1)$, we can apply Lemma 6.5 to obtain that whenever $g_P(t) = \max_Q g_Q(t)$ where $P = v_0, \ldots, v_k$, node $v_k$ is in slow mode and $\frac{d}{dt} L_{v_k}(t) = \frac{d}{dt} H_{v_k}(t) \leq 1 + \rho$. This is sufficient to apply Lemma 6.2, which yields (18). □

Intuitively, until now we argued that if $v_0$ is too far ahead of other nodes then those nodes will be in fast mode, and if $v_0$ is too far behind other nodes then those nodes will be in slow mode. What does $v_0$ *itself* do when it is too far behind? Observe that if there is some path $P = v_0, \ldots, v_k$ such that $\Psi_P^s(t) > 0$, then for the inverted path $P' = v_k, \ldots, v_0$ we have $\Xi_{P'}^s(t) > \Psi_P^s(t) > 0$. Thus, informally speaking, whenever $v_0$ is too far behind some other node it will be "pulled forward" at the fast rate. The next lemma quantifies how much ground $v_0$ makes up during an interval in which it is far behind: it states that given sufficient time, the node makes up all the initial weighted skew $\Psi_v^s$, *in addition* to its minimal rate of progress $(1 - \rho)$.

**Lemma 6.9.** *For any node $v_0$, integer $s \in \mathbb{N}^{>0}$ and time interval $[t_0, t_1]$ where $t_1 \geq t_0 + \frac{C_{s-1}}{(1-\rho)\mu}$, if the network is in a legal state at time $t_0$, then*

$$\mathcal{I}_{v_0}(t_0, t_1) \geq \Psi_{v_0}^s(t_0) + (1 - \rho)(t_1 - t_0).$$

*Proof.* If $\Psi_{v_0}^s(t_0) \leq 0$, the claim follows immediately from Property 6.6. Thus, assume that $\Psi_{v_0}^s(t_0) > 0$, and let $P = v_0, \ldots, v_k$ be a path such that $\Psi_P^s(t_0) = \Psi_{v_0}^s(t_0)$. From the definitions of $\Psi$ and $\Xi$, for the inverted path $P' = v_k, \ldots, v_0$ we have $\Xi_{P'}^s(t_0) > \Psi_P^s(t_0)$, and therefore, $\Xi_{v_k}^s(t_0) > \Psi_{v_0}^s(t) > 0$. If there is a time $t \in [t_0, t_1]$ such that $\Xi_{v_k}^s(t) \leq 0$, let $\bar{t}$ be the infimum of such times. Otherwise, let $\bar{t} = t_1$. Observe that

$$\begin{aligned}
\mathcal{I}_{v_0}(t_0, \bar{t}) &= L_{v_0}(\bar{t}) - L_{v_0}(t_0) = \\
&= (L_{v_k}(t_0) - L_{v_0}(t_0) - (s-1)\kappa_P) - (L_{v_k}(\bar{t}) - L_{v_0}(\bar{t}) - (s-1)\kappa_P) + L_{v_k}(\bar{t}) - L_{v_k}(t_0) = \\
&= \Xi_{P'}^s(t_0) - \Xi_{P'}^s(\bar{t}) + \mathcal{I}_{v_k}(t_0, \bar{t}) > \\
&> \Psi_P^s(t_0) - \Xi_{v_k}^s(\bar{t}) + \mathcal{I}_{v_k}(t_0, \bar{t}) = \\
&= \Psi_{v_0}^s(t_0) - \Xi_{v_k}^s(\bar{t}) + \mathcal{I}_{v_k}(t_0, \bar{t}).
\end{aligned}$$

Since $\bar{t} \leq t_1$ and $\mathcal{I}_{v_0}(t_0, \cdot)$ is non-decreasing and interval-additive, to prove the claim it is sufficient to show that $\mathcal{I}_{v_k}(t_0, \bar{t}) \geq \Xi_{v_k}^s(\bar{t}) + (1 - \rho)(\bar{t} - t_0)$.

Consider first the case where $\bar{t} < t_1$. In this case $\bar{t}$ is the infimum of times $t$ where $\Xi_{v_k}^s(t) \leq 0$. Since $\Xi_{v_k}^s(\cdot)$ is continuous, it follows that $\Xi_{v_k}^s(\bar{t}) = 0$, and using Property 6.6 we obtain $\mathcal{I}_{v_k}(t_0, \bar{t}) \geq \Xi_{v_k}^s(\bar{t}) + (1 - \rho)(\bar{t} - t_0)$.

Otherwise, if $\bar{t} = t_1$, then for all $t \in [t_0, t_1)$ we have $\Xi_{v_k}^s(t) > 0$. Applying Lemma 6.7 to the interval $[t_0, t_1]$ we obtain

$$\Xi^s_{v_k}(t_1) \leq \Xi^s_{v_k}(t_0) + \mathcal{I}_{v_k}(t_0, t_1) - (1-\rho)(1+\mu)(t_1 - t_0) \leq$$
$$\stackrel{\text{Lemma 6.3}}{\leq} C_{s-1} + \mathcal{I}_{v_k}(t_0, t_1) - (1-\rho)\mu \cdot \frac{C_{s-1}}{(1-\rho)\mu} - (1-\rho)(t_1 - t_0) =$$
$$= \mathcal{I}_{v_k}(t_0, t_1) - (1-\rho)(t_1 - t_0),$$

which yields the desired result. $\square$

Now we are ready to put all the pieces together and prove the main theorem:

**Theorem 6.10.** *The network is always in a legal state.*

*Proof.* Suppose for the sake of contradiction that this is not the case, and let $\bar{t}$ be the infimum of times when the legal state condition is violated. Then there is some path $P = v_0, \ldots, v_k$ and some $s \geq 1$ such that $\kappa_P \geq C_s$ but

$$L_{v_0}(\bar{t}) - L_{v_k}(\bar{t}) \geq s \cdot \kappa_P. \tag{20}$$

For the legal state condition to be violated, the system must be far outside the boundary of $\mathbf{\Psi}^s$:

$$\Psi^s_{v_k}(\bar{t}) \geq L_{v_0}(\bar{t}) - L_{v_k}(\bar{t}) - \left(s - \frac{1}{2}\right) \cdot \kappa_P \stackrel{(20)}{\geq} \frac{1}{2}\kappa_P \geq \frac{1}{2}C_s = \frac{1}{2\sigma}C_{s-1}. \tag{21}$$

However, Lemma 6.8 tells us that whenever $\Psi^s_{v_k}$ is large it cannot increase quickly, which gives $v_k$ time to catch up. More specifically, if $t_0$ is the supremum of times $t \leq \bar{t}$ such that $\Psi^s_{v_k}(t) \leq 0$, then Lemma 6.8 shows that

$$\Psi^s_{v_k}(\bar{t}) \leq \Psi^s_{v_k}(t_0) - \mathcal{I}_{v_k}(t_0, \bar{t}) + (1+\rho)(\bar{t} - t_0) \stackrel{\text{(Prop. 6.6)}}{\leq} 2\rho(\bar{t} - t_0), \tag{22}$$

and combining (21) and (22) we see that $t_0 \leq \bar{t} - \frac{C_{s-1}}{4\sigma\rho} \stackrel{(3)}{\leq} \bar{t} - \frac{C_{s-1}}{(1-\rho)\mu}$. According to Lemma 6.9, this is sufficient time for $v_k$ to increase its clock by

$$\mathcal{I}_{v_k}(t_0, \bar{t}) \geq \Psi^s_{v_k}(t_0) + (1-\rho)(\bar{t} - t_0), \tag{23}$$

which we combine with the first inequality of (22) to obtain

$$\Psi^s_{v_k}(\bar{t}) \stackrel{(22)}{\leq} \Psi^s_{v_k}(t_0) - \mathcal{I}_{v_k}(t_0, \bar{t}) + (1+\rho)(\bar{t} - t_0) \stackrel{(23)}{\leq} 2\rho\frac{C_{s-1}}{(1-\rho)\mu} \stackrel{(3)}{<} \frac{1}{2\sigma}C_{s-1},$$

in contradiction to (21). $\square$

As an easy corollary we obtain the following.

**Theorem 6.11.** *The global skew of the algorithm is bounded by $2\mathcal{D}$.*

*Theorem 6.11.* Theorem 6.10 allows us to use Lemma 6.3 at any time $t$. For any two nodes $u, v$, let $P$ be a path from $u$ to $v$. Lemma 6.3 states (in particular) that

$$C_0 > \Xi_u^1(t) \geq \Xi_P^1(t) \geq L_v(t) - L_u(t) - (1-1)\kappa_P = L_v(t) - L_u(t),$$

and since $C_0 = 2\mathcal{D}$, the claim follows. [1] □

**Corollary 6.12.** *If $\sigma = \Theta(1/\rho)$, $\mu = \Theta(1/(1-\rho))$ and $\kappa_e = \Theta(\epsilon_e)$ for all $e \in E^{\text{est}}$, then the algorithm achieves $O\left(\text{dist}(u,v) \cdot \log_{1/\rho} \mathcal{D}\right)$-gradient synchronization, with a global skew of $O(\mathcal{D})$.*

---

[1] We note that the proof of Theorem 6.10 goes through with any value of $C_0 > \mathcal{D}$, which allows to improve the global skew bound to $(1 + \epsilon)$ for arbitrarily small $\epsilon$. This bound is asymptotically optimal, but the constant can be improved further by having nodes keep track of the maximal clock in the network, as was done in [12].

# Appendix

## A The Estimate Layer

### A.1 Error analysis for direct estimates

To analyze the error in a direct estimate at time $t$, let $t_{\text{snd}}$ be the last time node $v$ sends a message that node $u$ receives by time $t$. Let $t_{\text{rcv}} \leq t$ be the time when $u$ receives the message, and let $L$ be the clock value the message carries.

During the interval $[t_{\text{rcv}}, t]$, node $u$ increases $\tilde{L}_u^{v,\text{direct}}$ at the rate of its own hardware clock, and therefore

$$(1 - \rho)(t - t_{\text{rcv}}) \leq \tilde{L}_u^{v,\text{direct}}(t) - L \leq (1 + \rho)(t - t_{\text{rcv}}). \tag{24}$$

Also, from Requirement 3.2,

$$(1 - \alpha)(t - t_{\text{snd}}) \leq L_v(t) - L \leq (1 + \beta)(t - t_{\text{snd}}). \tag{25}$$

Because $t_{\text{rcv}} \in [t_{\text{snd}}, t_{\text{snd}} + \mathcal{T}]$, we can re-write (24) to obtain

$$(1 - \rho)(t - t_{\text{snd}} - \mathcal{T}) \leq \tilde{L}_u^{v,\text{direct}}(t) - L \leq (1 + \rho)(t - t_{\text{snd}}), \tag{26}$$

and subtracting (26) from (25) yields

$$-(\alpha + \rho)(t - t_{\text{snd}}) \leq L_v(t) - \tilde{L}_u^{v,\text{direct}}(t) \leq (\beta + \rho)(t - t_{\text{snd}}) + (1 - \rho)\mathcal{T}. \tag{27}$$

Finally, since $v$ broadcasts every $\Delta H$ subjective time units, at time $t'_{\text{snd}} \leq t_{\text{snd}} + \frac{\Delta H}{1 - \rho}$ node $v$ broadcasts again, and the second broadcast is received by $u$ at time $t'_{\text{snd}} + \mathcal{T}$ at the latest. The second broadcast is not received by time $t$, and it follows that $t_{\text{snd}} \geq t - \frac{\Delta H}{1 - \rho} - \mathcal{T}$. Substituting this bound in (27), we get

$$-(\alpha + \rho)\left(\frac{\Delta H}{1 - \rho} + \mathcal{T}\right) \leq L_v(t) - \tilde{L}_u^{v,\text{direct}}(t) \leq (\beta + \rho)\left(\frac{\Delta H}{1 - \rho} + \mathcal{T}\right) + (1 - \rho)\mathcal{T}.$$

### A.2 Error analysis for RBS estimates

Let $v \in N^2(u)$. Suppose that at time $t$, time $t_x$ is the latest time an event $x$ occurs such that node $u$ receives a report$(v, x, L)$ message by time $t$. Let $t_{\text{rcv}}$ be the time at which $u$ receives the report, and let $t_x^v$ be a time such that $L = L_v(t_x^v)$, and let $t_x^u$ be a time such that $H = H_u(t_x^u)$. We know that $t_x \leq t_x^u, t_x^v \leq t_x + u_{\text{rcv}}$.

As before, we have

$$(1 - \rho)(t - t_{\text{rcv}}) \leq \tilde{L}_u^{v,\text{rbs}}(t) - L - H_u(t_{\text{rcv}}) + H_u(t_x^u) \leq (1 + \rho)(t - t_{\text{rcv}}), \tag{28}$$

$$(1 - \rho)(t_{\text{rcv}} - t_x^u) \leq H_u(t_{\text{rcv}}) - H_u(t_x^u) \leq (1 + \rho)(t_{\text{rcv}} - t_x^u), \tag{29}$$

and
$$(1 - \alpha)(t - t_x^v) \leq L_v(t) - L \leq (1 + \beta)(t - t_x^v). \tag{30}$$

Summing (28) and (29) yields
$$(1 - \rho)(t - t_x^u) \leq \tilde{L}_u^{v,\text{rbs}}(t) - L \leq (1 + \rho)(t - t_x^u), \tag{31}$$

and because $t_x^u, t_x^v \in [t_x, t_x + u_{\text{rcv}}]$, we can re-write (30) and (31) as
$$(1 - \rho)(t - t_x - u_{\text{rcv}}) \leq \tilde{L}_u^{v,\text{rbs}}(t) - L \leq (1 + \rho)(t - t_x) \tag{32}$$

and
$$(1 - \alpha)(t - t_x - u_{\text{rcv}}) \leq L_v(t) - L \leq (1 + \beta)(t - t_x). \tag{33}$$

Subtracting (32) from (33) we obtain
$$-(\alpha + \rho)(t - t_x) - (1 - \alpha)u_{\text{rcv}} \leq L_v(t) - \tilde{L}_u^{v,\text{rbs}}(t) \leq (\beta + \rho)(t - t_x) + (1 - \rho)u_{\text{rcv}}. \tag{34}$$

Since every node broadcasts every $\frac{\Delta H}{1-\rho}$ time units at most, at some time $t_y \leq t_x + \frac{\Delta H}{1-\rho}$ the common neighbor of $u$ and $v$ will broadcast again, and both nodes will record the event. The corresponding report$(\cdot)$ will be received by $u$ no later than time $t_y + \mathcal{P}$. Since no such message is received before time $t$, we have $t_x \geq t - \frac{\Delta H}{1-\rho} - \mathcal{P}$. Substituting in (34), we get

$$-(\alpha + \rho)\left(\frac{\Delta H}{1-\rho} + \mathcal{P}\right) - (1 - \alpha)u_{\text{rcv}} \leq L_v(t) - \tilde{L}_u^{v,\text{rbs}}(t) \leq$$
$$\leq (\beta + \rho)\left(\frac{\Delta H}{1-\rho} + \mathcal{P}\right) + (1 - \rho)u_{\text{rcv}}.$$